**This document is an unpublished manuscript.**

# Validation of Sinus Drug Delivery Computational Fluid Dynamics (CFD) Modeling with In Vitro Gamma Scintigraphy

Kathryn Kudlaty MD[1], William Bennett Ph.D[2], Landon Holbrook Ph.D[2], Alyssa Burke BS[2], Jihong Wu MD[2], Benjamin Langworthy BS[3], Jason P. Fine Ph.D[3], Charles S. Ebert, Jr. MD[1], Adam J. Kimple MD, Ph.D[1], Brian D. Thorp MD[1], Adam M. Zanation MD[1], Brent A. Senior MD[1], Julia S. Kimbell Ph.D[1].

[1]Department of Otolaryngology, Head & Neck Surgery, University of North Carolina, Chapel Hill

[2]Center for Environmental Medicine, Asthma and Lung Biology, University of North Carolina

[3]Department of Biostatistics, University of North Carolina, Chapel Hill

**Correspondence:** Kathryn Kudlaty

UNC Hospitals - Department of Otolaryngology, Head & Neck Surgery

170 Manning Drive, CB 7070

Physicians Office Building, Rm G190A

Chapel Hill, NC  27599-7070

Phone: 817-247-3665

Fax:    919-966-7941

Email: kathryn.kudlaty@unchealth.unc.edu

**Header Title:** CFD Validation for Sinus Drug Delivery

**Disclosures:** Research reported in this publication was supported by the National Heart, Lung and Blood Institute of the National Centers of Health under award number R01HL122154.  The content is solely the responsibility of the authors and does not necessarily represent the official views of the National Institutes of Health. The authors have no financial disclosures or conflicts of interest to report.

**ABSTRACT**

**Background**: Chronic rhinosinusitis (CRS) is a prevalent and disruptive disease. Medical management including nasal steroid sprays is the primary treatment modality. Computational fluid dynamics (CFD) has been used to characterize sinonasal airflow and intranasal drug delivery; however, variation in simulation methods indicates a need for large scale CFD model validation.

   **Methods**: Anatomic reconstructions of pre and post-operative CT scans of 3 functional endoscopic sinus surgery patients were created in Mimics(TM). Fluid analysis and drug particle deposition modeling were conducted using CFD methods with Fluent(TM) in 18 cases. Models were 3D printed and in vitro studies were performed using Tc99-labeled Nasacort(TM). Gamma scintigraphy signals and CFD-modeled spray mass were post-processed in a superimposed grid and compared. Statistical analysis using overlap coefficients (OCs) evaluated similarities between computational and experimental distributions and Kendall's tau rank correlation coefficient was employed to test independence.

   **Results**: OCs revealed strong agreement in percent deposition and grid profiles between CFD models and experimental results (mean [range] for sagittal, axial, and coronal grids were 0.69 [0.57], 0.61 [0.49], and 0.78 [0.44], respectively). Kendall's tau values showed strong agreement (average 0.73) between distributions, which were statistically significant ($p < 0.05$) apart from a single coronal grid in one model and two sagittal grids of another.

   **Conclusions**: CFD modeling demonstrates statistical agreement with in vitro experimental results. This validation study is one of the largest of its kind and supports the applicability of CFD in accurately modeling nasal spray drug delivery and using computational methods to investigate means of improving clinical drug delivery.

**Introduction:**

Chronic rhinosinusitis (CRS) is an inflammatory condition characterized by the cardinal symptoms of: nasal congestion, nasal drainage, facial pain and pressure, and decreased smell persisting for longer than 12 weeks[1]. It is a profoundly prevalent disease, affecting approximately 12% of the US population and an estimated 5-12% worldwide[2,3]. This results in similarly heavy financial and emotional burdens on both an individual and population level. Annual spending for CRS in the US was projected to be up to $10.8 billion dollars by 2020 and is still increasing[4]. Indirect costs considering loss of productivity and psychologic burden are even higher, with self-reported perceptions of health (measured by health utility scores) by CRS patients comparable to patients with Parkinson's disease and end stage renal disease with dialysis dependence[2].

The treatment of CRS is largely medical. In addition to nasal saline irrigation, nasal corticosteroid sprays are the gold standard of initial management and function to modulate the underlying inflammatory mediators of disease[1]. Despite widespread use, approximately 50% of patients fail appropriate medical treatment[5,6]. This is attributed in part to the limited penetration of medication within the sinonasal passage, a limitation which persists even after surgical intervention[7,8]. Several factors have been identified in contributing to suboptimal deposition at target sites, including: nozzle position, inhalation characteristics, and particle size[9-13].

In an effort to improve drug delivery and thus outcomes to medical management, increasing study has focused on characterizing the location and amount of drug delivery achieved using current administration methods. Computational fluid dynamics (CFD) has arisen in recent decades as an effective tool for modeling sinonasal airflow and its role in intranasal drug delivery[10,11,14,15]. In addition to overcoming the logistical and ethical burdens of measuring drug delivery in vivo, CFD modeling offers the immediate ability to measure the effect of an alteration in medication administration, such as particle size or position of delivery device, in a way that lends itself well to optimizing medical intervention.

A number of previous studies have implemented CFD to characterize sinonasal airflow and intranasal drug delivery[10,14-18]. Since all CFD studies are subject to a number of simplifying assumptions, validation of CFD results is necessary[19]. The validation studies that have been performed comparing in vitro with in silico modeling are largely limited by their relatively small scale, often isolated to a single anatomical model, and by the broad manner in which drug deposition location is characterized[20,21]. Deposition location is predominantly described in either large subunits (anterior/middle/posterior compartments) or by percent penetration past a particular anatomic landmark (i.e. internal nasal valve), limiting the statistical comparison between in vitro and in silico models[8,9,20,21]. A recent CFD investigation of deposition increases achieved by aiming a nasal sprayer toward specific nasal target sites in 3 subjects with CRS used gamma scintigraphy in both nasal sides of one 3D-printed model to confirm simulated spray mass[22]. The study presented here expands on this work to provide the largest scale validation of its kind to date, using overlap coefficients to compare CFD modeling and experimental results in 18 distinct cases with varying sets of spray use conditions and pre vs post-surgical changes. This work supports the use of CFD methods to explore and quantify nasal drug delivery improvements.

**Materials and Methods:**

*Computational model creation:*

To generate the models for CFD analysis, anatomic reconstructions were made of de-identified CT scans previously obtained as part of an ongoing study protocol and approved by the institutional review board (IRB) at the University of North Carolina at Chapel Hill. Pre- and post-operative CT scans were obtained for 3 functional endoscopic sinus surgery (FESS) patients (Table 1).

Anatomical reconstructions were created in Mimics™ 18.0 imaging software (Materialise, Inc., Plymouth, MI, US) with initial Hounsfield units (HU) thresholding values from -1024 to -300 (Siemens) and -1024 to 300 or 400 (cone-beam CT, CBCT) used to designate the airspace of the sinonasal passages and paranasal sinuses. Reconstructed airspaces were then hand-edited for accuracy and reviewed on a case-by-case basis by a clinical rhinologist. These models were then imported into ICEM-CFD™ 15.0 (ANSYS, Inc., Canonsburg, PA, US) as previously described[15,23]. Inlet and outlet boundary surfaces were designated at the nostril inlets and nasopharynx, respectively. For analytical comparison purposes with the experimental results, a 3D grid was designed in ICEM™ and superimposed on the model based on designated reference points (Figure 1). Grids comprised planes in the coronal (xy), sagittal (yz), and axial (xz) orientations and were designed to match the grid designation created in the experimental analysis, described below.

Tetrahedral meshes of at least 3.8 million cells with quality > 0.3 were also created in ICEM™, in a manner described in previous work[16]. Three 0.1-mm-thick prism layers were then added and the final, hybrid meshes were smoothed globally until the number of elements in the 0 – 0.1 quality range was smaller than 0.0005%.

*Fluid modeling and drug deposition:*

Meshes were then exported to Fluent™ for modeling of airflow and drug particle deposition. Flow parameters assumed laminar nasal airflow, as supported by previous findings[24-26]. To mimic the gentle inspiration directed by nasal spray use instructions, steady inspiratory airflow simulations were carried out at twice the resting minute volume using methods described previously[23,27]. Resting breathing minute volume for each patient was measured pre-operatively using LifeShirt plethysmography (range 9 to 12 L/min)[28]. In the simulations, gauge pressure was set at 0 at the nostrils with a negative pressure set at the nasopharynx sufficient to generate the patient-specific flow rate (range -7.4 to -18.9 Pa). Fluent™ provides numerical solutions for flow characteristics by solving the differential equations governing conservation of mass and momentum for laminar, incompressible flow as described in detail in previous work[23,27]. Acceptable convergence was determined by small residuals and stabilization of outlet mass flow rate.

Following fluid analysis, drug particle deposition was determined through Fluent's™ Discrete Phase Model (DPM) with Langrangian particle tracking performed by integrating the particle transport equation of the Runge-Kutta method[20]. Initial parameters were defined in such a way as to introduce droplet particles into the system in a manner consistent with nasal spray actuation. These parameters were based on studies of particle size and plume geometry of Triamcinolone acetonide (Nasacort™) nasal sprays performed by Next Breath, LLC as described previously, resulting in a volume-based droplet size distribution with a Dv50 of 43.81 μm and GSD of 1.994 μm and cone half angle of 27.93°.[22] Spray velocity was set to 18.5 m/s.[29] Drug particle size distribution was characterized using the formula described by Cheng et al[30]. As our validation study was concerned with comparing fractional deposition location rather than absolute amount of drug delivery, this size distribution was scaled to a 1mg total delivered amount to reduce the number of particles and alleviate the time burden of CFD modeling. Sprays were simulated as release from a point at the tip of the virtual nozzle in either the left or right nasal vestibule. As the presence of the nozzle has been previously shown to be largely negligible, it was not included within the 3D space of the model[23].

Airflow was solved for each of the 6 anatomical models (pre- and post-operative for the 3 patients, models labeled "PRE" and "POST", respectively). As each model provided distinct left and right-hand sides (labeled "LHS" and "RHS"), there were a total of 12 regions of interest. Additionally, nasal drug delivery was simulated with the nozzle in two different positions: "current use" (CU), consistent with provided medication instructions, or by "line of sight" (LOS), in which the nozzle was optimally orientated toward the ostiomeatal complex[22]. Although outside the scope of this validation study, evaluating LOS as a manner of improving nasal drug delivery is a topic of ongoing interest[22]. This set of conditions, sides, and models resulted in the potential for 24 distinct trials for comparison purposes. This was reduced to a total of 18 due to anatomical constraints (certain models lacked a clear "line of sight") and a lack of experimental data for the right-hand side of model SD02.

*Experimental design:*

As described in previous work, a 3D printer was used to print the pre- and post-surgical models for the three study subjects[23]. The anterior portion of the model comprising the external nares and anterior vestibules was printed from a flexible component which was fitted on to the posterior portion of the model, which was printed from a rigid Watershed material (DSM Somos, Elgin, Illinois). This allowed for easy removal of the anterior portion of the model to isolate the signal depositing within the regions of interest deeper within the sinonasal passages. Aiming devices were also 3D printed to interface with the medication nozzle and the external nare to ensure consistency in nozzle position with CFD modeling and between individual experimental trials. Experimental set-up is the same as that established in previous work by this group[22] (Figure 2).

To measure amounts and location of drug deposition experimentally, a small solution of $^{99m}$Technetium as sodium pertechnetate, Na[99mTcO4] was added to the medication for a single actuation of no more than 10 µCi of activity as in published methods[31,32]. A steady inspiratory flow of air was drawn through the replica to simulate gentle inspiration on a patient-specific basis[27,31]. A filter was placed to collect any particles that traveled to the nasopharyngeal outlet and a tissue used to retrieve any residual deposited spray dripping from the external nose to be included in the total signal measurement. Medication was hand actuated a single time into one nostril in either current use (CU) or line of sight (LOS) orientations. A medium resolution large field of view gamma camera (Body Scan, MiE America, Inc., Elk Grove Village, IL) was used to measure $^{99m}$Tc gamma activity associated with the deposition pattern for the mass of labeled spray within the model. Resolution of the gamma scans was determined by pixel size, with one pixel corresponding to 2.38mm for a 256 x 256 matrix scan. Using this unit of measurement, a grid system was created to assign gamma activity of drug deposition to compartments for comparison purposes (Figure 3). Grid lines were spaced 4 pixels (9.52mm) apart in sagittal and axial orientations. Given the predominant spray deposition at the medial aspect of the model, grid lines were spaced closer together at the midline of the model in the frontal view to provide increased specificity (range from 2.38 – 14.28mm).

*Post-processing:*

Solved particle trajectories in Fluent$^{TM}$ provided Cartesian coordinates of individual particle deposition locations that were collected in Microsoft® Excel. Data was post-processed to exclude points landing within the anterior portion of the nose. Particles were assigned to grid compartments by location and associated masses were summed to determine total deposited dose per each compartment. Since CFD modeling was performed using a scaled down spray of representative particle mass distribution, as described above, mass amounts were converted to percent mass in order to perform statistical comparison with experimental results.

Despite matching grids to pre-defined reference points (americium markers) and using the established grid spacing, the virtual grid in ICEM$^{TM}$ did not exactly match that applied to the experimental model. There were several sources identified in contributing to this effect, the first being that the experimental grid was limited by the resolution of the gamma-scintigraphy camera, with smallest discrete pixel sizes measuring 2.38mm. The dimensions of the physical model and americium marker spacing were not designed in units of pixel size and thus did not fall along exact increments. Due to the nature of gamma-scintigraphy signal capturing and image processing, grid positioning was performed manually, with small shifts made to maximize americium signals within identified marker locations. This was further complicated by the effect of visual perspective and foreshortening in captured images that was absent from the computational models. To allow for this reasonable variation in grid positioning, a "grid shift" was applied to the virtual grid system, in which each orthogonal set of planes was linearly translated 1 pixel distance (2.38mm) in each direction. The original plane positions are heretofore referred to as "reference planes" with the shifted planes designated as "positive shift" and "negative shift" based on direction of translation along the corresponding axes (Figure 4).

*Statistical analysis:*

Statistical analysis was performed using overlap coefficients (OCs) to evaluate similarities between computational and experimental distributions. This measures the proportion of overlap between the two deposition distributions. Values range between 0 and 1, with a value of 0 indicating no overlap and a value of 1 representing a perfect match. This was performed separately for distributions in each orthogonal set of planes, as represented by the equation:

$$OC = 1 - \frac{\sum_{i=1}^{n}|sim_i - exp_i|}{2}$$

where n is the number of grid compartments along each corresponding axis, $sim_i$ represents the proportion of simulation deposition in the $i^{th}$ grid compartment along the axis, and $exp_i$ is the corresponding experimental deposition in that grid compartment.

In addition, the Kendall's tau rank correlation coefficient and the p-value for the Kendall's tau test for independence were obtained. The p-values for the test for independence were false discovery rate corrected using the Benjamini-Hochberg procedure. The correction was based on all 18 models for all three directions and the reference, negative, and positive grid positions.

**Results:**

For each model, fractional mass deposition was post-processed into grid compartments along each axis for both computational and experimental data. Analysis was limited to deposition within the posterior rigid portion of the model (excluding the external nose and anterior vestibule) in the anatomical region of interest. In CFD models, a single simulated spray actuation resulted in percent depositions ranging from 2.07% for Model SD04PRE-LHS-CU to 63.63% for Model SD05PRE-RHS-LOS with a mean percent deposition of 27.99%. The percent deposition in experimental trials ranged from 16.43% for Model SD05POST-RHS-CU to 57.66% for Model SD02POST-LHS-LOS with a mean of 34.74%. Due to the inherent dependence between grid compartments, with particle masses falling into one compartment or another, graphs were generated to more easily depict general trends in mass deposition location within each model. A representative sample is shown below (Figure 5) illustrating percent mass deposition within computational and experimental compartments for Model SD05PRE-LHS-LOS. The computational results include the reference grids as well as the positive and negative shifts. Experimental results are represented by mean value across multiple trials with associated error bars.

In order to compare the computational and experimental data to evaluate the accuracy of CFD modeling it was necessary to analyze overall distribution among grid compartments. This was performed using overlap coefficients (OCs), which are provided for each model by grid system in Tables 2-4. Mean OCs were highest in coronal plane compartments, followed by sagittal and then axial distributions.

The OCs among coronal plane compartments were highest in the reference grid position (mean 0.78) compared to negative and positive shifts (0.77 and 0.73, respectively). The highest OCs for sagittal and axial planes were achieved with the positively-shifted plane positions; however the reference position never provided the worst agreement. For sagittal planes, mean OCs in descending order were 0.78 (positive shift), 0.69 (reference planes), and 0.51 (negative shift). For axial planes mean OCs were 0.63 (positive shift), 0.61 (reference planes), and 0.54 (negative shift).

Also provided in these tables are the associated Kendall's tau rank correlation coefficients, which were used to assess independence between computational and experimental data, with their representative p-values. Mean Kendall's tau values of distributions in reference plane positions were 0.66 (coronal), 0.76 (sagittal), and 0.76 (axial). P-values were all significant ($p \leq 0.05$) apart from two sets of sagittal grids for one model (SD04POST-RHS-CU) and one shifted coronal grid for another (SD05PRE-RHS-LOS).

**Discussion:**

The results of this study reveal strong agreement in percent deposition and grid profiles between CFD models and experimental results as measured by OCs. This method of validation is novel in the statistical methods of rank correlation as well as the independent evaluation of drug deposition along each axis and the use of small geometric compartments rather than large subunits for comparison purposes. Although limited in direct anatomic correlates, reporting mass deposition by grid compartments is still effective in demonstrating depth of particle penetration within the sinonasal passages and better serves the underlying objective of providing numerous, uniform bins for increased sensitivity in statistical analysis.

      There were certain physical limitations inherent in the experimental model. The foremost among these was the relatively large pixel size resulting from the resolution of the gamma scintigraphy images and the difficulty in creating a uniform grid to accurately overlie model geometries that were not designed with this unit of measurement in mind. Although shifting the best-estimate reference grids was performed to evaluate the most accurate grid position, this effect hinders perfect concordance between the constructed systems of comparison. The consistently improved OC values for the positively-shifted sagittal and axial planes suggests a possible underlying skew in the manual adjustments performed in superimposing the grid over the gamma scintigraphy results, a source of error that may be minimized in the future by designing model reference points with these considerations in mind. Multiple experimental trials were performed for each model and set of spray conditions; however, this was limited by the time-intensive nature of radiotracer use in a given model. Another potential source of variation among experimental trials was the hand actuation of the nasal spray. Although a dedicated aimer was implemented to provide consistency in positioning, hand actuation was performed by different individuals with unavoidable variations in applied pressure and performance. Few cases, especially those with low amounts of particle deposition, demonstrated a marked increase in fractional experimental deposition relative to computational results. This is perhaps due to the beneficial effect of mechanical stenting of the nasal valve resulting from nozzle deformation in the anterior nose. These cases still demonstrated strong overlap coefficients, signifying overall agreement between drug deposition distributions.

Limitations in CFD modeling arise from certain assumptions applied in simplifying the simulation. These include releasing spray particles from the nozzle at a single point source as well as having particles instantly assume the volume distribution pattern measured experimentally at a distance 3cm from the nozzle tip. Models could further be refined by considering the effects of humidity within the sinonasal passages and the corresponding alterations in particle behavior and volume.

In using OCs to evaluate the correlation between dependent bins, there is an additional degree of concordance not necessarily represented numerically. This derives from the fact that compartment bins that had discrepancies in amounts of overlap between CFD and experimental models were often located adjacent to one another. This is more easily appreciated graphically, as illustrated in Figure 6. This reflects similarity in the overall profile of mass distributions, rather than discrepancies arising from compartments in anatomically unrelated locations.

Mean OCs were highest in coronal plane compartments, representing best overall agreement in modeling drug deposition in an anterior to posterior direction. This further supports the clinical strength of this modeling system, as this distribution represents the depth of penetrance of drug particles past the internal nasal valve into the sinonasal passage and has been the focus of much of the existing work in CFD modeling of intranasal drug delivery[22,32].

The additional use of Kendall's rank correlation coefficients provided a less robust, but additionally useful measure of independence between computational and experimental results. Although these values supported independence for the vast majority of models, this was not proven for the sagittal compartments in the reference and negative shift positions for Model SD04POST-RHS-CU. This is attributed to the fact that in this axis the positively shifted grid is demonstrated to provide a better correlation and is likely the most accurate grid position. Similarly, the negatively shifted coronal grid in SD05PRE-RHS-LOS demonstrated a lack of significance, but the reference plane position resulted in a strong OC and Kendall's tau value suggesting a more accurate fit in this case.

Although mean fractional deposition in the posterior part of the model was similar between CFD and experimental models (27.99% and 34.74%, respectively), these values reveal that the majority of medication is deposited in the anterior portion of the nose, a limitation that has been established in the literature and largely attributed to the obstructive nature of the internal nasal valve[9,30,31,33]. Although outside the direct scope of this validation study, this trend is easily identified in the collected data. While previous CFD studies have found negligible effects from modeling space occupied by a spray nozzle in the nostril, the effects of possible mechanical alterations to the internal nasal framework have not been examined[23]. With the establishment of CFD as an effective tool for evaluating nasal drug delivery, it is hoped that the factors contributing to this overall poor drug delivery may be identified and remedied.

**Conclusion:**

The described method of CFD modeling demonstrates statistical agreement with in vitro experimental results. This validation study is one of the largest of its kind and supports the applicability of CFD in accurately modeling nasal spray drug delivery. Computational models facilitate investigations into methods of improving clinical drug delivery without the associated financial, physical, and time burdens of physical experimentation. Future work is ongoing in determining exactly which methods of spray administration are successful in achieving optimal drug delivery and the additional effects of surgical intervention on improving such delivery.

**Acknowledgments:**

Reported research was supported by the National Heart, Lung, and Blood Institute of the National Institutes of Health (NIH), under award number R01HL122154. The content is solely the responsibility of the authors and does not necessarily represent the official views of the NIH. The authors thank Cara Breeden, BS and Julie D. Suman, PhD at Next Breath, LLC for furnishing the experimental findings on spray droplet sizes and plume geometries.

Kudlaty-17

**TABLES**

| Subject | Patient Characteristics | Surgical Procedures |
|---------|------------------------|---------------------|
| **SD02** | 70-year old<br>Male, 67.5kg<br>Caucasian | Bilateral FESS |
| **SD04** | 24-year old<br>Female, 93.1kg<br>Caucasian | Septoplasty<br>Bilateral maxillary antrostomy,<br>Bilateral anterior ethmoidectomy,<br>Bilateral inferior turbinate resection |
| **SD05** | 41-year old<br>Male, 88kg<br>Caucasian | Septoplasty<br>Bilateral FESS |

Table 1 – Study subject demographics and performed surgical procedures. FESS here refers to comprehensive surgery of the maxillary, ethmoid, sphenoid, and frontal sinuses.

| CORONAL PLANES | Overlap Coefficients (OCs) | | | Kendall's tau values | | | p-values | | |
|---|---|---|---|---|---|---|---|---|---|
| **Model** | Neg | **Ref** | Pos | Neg | **Ref** | Pos | Neg | **Ref** | Pos |
| SD02POST-LHS-CU | 0.84 | **0.6** | 0.59 | 0.77 | **0.75** | 0.75 | 0.00 | **0.00** | 0.00 |
| SD02POST-LHS-LOS | 0.81 | **0.83** | 0.82 | 0.77 | **0.8** | 0.77 | 0.00 | **0.00** | 0.00 |
| SD02PRE-LHS-CU | 0.87 | **0.78** | 0.71 | 0.80 | **0.75** | 0.75 | 0.00 | **0.00** | 0.00 |
| SD02PRE-LHS-LOS | 0.78 | **0.85** | 0.85 | 0.78 | **0.81** | 0.80 | 0.00 | **0.00** | 0.00 |
| SD04POST-LHS-CU | 0.81 | **0.64** | 0.56 | 0.69 | **0.64** | 0.69 | 0.00 | **0.00** | 0.00 |
| SD04POST-LHS-LOS | 0.73 | **0.74** | 0.69 | 0.67 | **0.69** | 0.69 | 0.00 | **0.00** | 0.00 |
| SD04POST-RHS-CU | 0.43 | **0.47** | 0.51 | 0.43 | **0.51** | 0.59 | 0.04 | **0.02** | 0.01 |
| SD04PRE-LHS-CU | 0.61 | **0.72** | 0.57 | 0.46 | **0.51** | 0.59 | 0.03 | **0.02** | 0.01 |
| SD04PRE-LHS-LOS | 0.9 | **0.83** | 0.74 | 0.64 | **0.61** | 0.61 | 0.00 | **0.00** | 0.00 |
| SD04PRE-RHS-CU | 0.89 | **0.79** | 0.72 | 0.67 | **0.64** | 0.79 | 0.00 | **0.00** | 0.00 |
| SD04PRE-RHS-LOS | 0.72 | **0.86** | 0.86 | 0.61 | **0.56** | 0.64 | 0.00 | **0.01** | 0.00 |
| SD05POST-LHS-CU | 0.86 | **0.9** | 0.75 | 0.70 | **0.75** | 0.69 | 0.00 | **0.00** | 0.00 |
| SD05POST-LHS-LOS | 0.7 | **0.75** | 0.8 | 0.55 | **0.66** | 0.72 | 0.02 | **0.00** | 0.00 |
| SD05POST-RHS-CU | 0.72 | **0.82** | 0.69 | 0.45 | **0.5** | 0.50 | 0.05 | **0.02** | 0.02 |
| SD05PRE-LHS-CU | 0.75 | **0.89** | 0.82 | 0.55 | **0.66** | 0.63 | 0.02 | **0.00** | 0.01 |
| SD05PRE-LHS-LOS | 0.81 | **0.91** | 0.86 | 0.64 | **0.78** | 0.72 | 0.00 | **0.00** | 0.00 |
| SD05PRE-RHS-CU | 0.71 | **0.82** | 0.81 | 0.61 | **0.55** | 0.47 | 0.01 | **0.02** | 0.03 |
| SD05PRE-RHS-LOS | 0.85 | **0.83** | 0.83 | 0.41 | **0.64** | 0.72 | 0.06 | **0.00** | 0.00 |
| **Mean** | **0.77** | **0.78** | **0.73** | **0.62** | **0.66** | **0.67** | | | |

Table 2 – Coronal planes. Overlap coefficients and Kendall's tau rank correlation coefficients with associated p-values demonstrating level of statistical agreement between mass deposition distributions analyzed by coronal (XY) planes

| SAGITTAL PLANES | Overlap Coefficients (OCs) | | | Kendall's tau values | | | p-values | | |
|---|---|---|---|---|---|---|---|---|---|
| **Model** | Neg | **Ref** | Pos | Neg | **Ref** | Pos | Neg | **Ref** | Pos |
| SD02POST-LHS-CU | 0.5 | **0.82** | 0.75 | 0.73 | **0.66** | 0.58 | 0.00 | **0.00** | 0.01 |
| SD02POST-LHS-LOS | 0.67 | **0.72** | 0.84 | 0.80 | **0.77** | 0.77 | 0.00 | **0.00** | 0.00 |
| SD02PRE-LHS-CU | 0.38 | **0.73** | 0.76 | 0.69 | **0.77** | 0.77 | 0.00 | **0.00** | 0.00 |
| SD02PRE-LHS-LOS | 0.7 | **0.84** | 0.8 | 0.80 | **0.80** | 0.75 | 0.00 | **0.00** | 0.00 |
| SD04POST-LHS-CU | 0.64 | **0.77** | 0.95 | 0.80 | **0.85** | 0.88 | 0.00 | **0.00** | 0.00 |
| SD04POST-LHS-LOS | 0.58 | **0.73** | 0.84 | 0.74 | **0.85** | 0.77 | 0.00 | **0.00** | 0.00 |
| SD04POST-RHS-CU | 0.25 | **0.27** | 0.56 | 0.19 | **0.27** | 0.43 | 0.39 | **0.21** | 0.05 |
| SD04PRE-LHS-CU | 0.47 | **0.82** | 0.92 | 0.73 | **0.77** | 0.70 | 0.00 | **0.00** | 0.00 |
| SD04PRE-LHS-LOS | 0.44 | **0.61** | 0.74 | 0.76 | **0.79** | 0.76 | 0.00 | **0.00** | 0.00 |
| SD04PRE-RHS-CU | 0.27 | **0.51** | 0.92 | 0.56 | **0.67** | 0.73 | 0.01 | **0.00** | 0.00 |
| SD04PRE-RHS-LOS | 0.45 | **0.66** | 0.91 | 0.73 | **0.78** | 0.77 | 0.00 | **0.00** | 0.00 |
| SD05POST-LHS-CU | 0.19 | **0.7** | 0.83 | 0.65 | **0.87** | 0.82 | 0.02 | **0.00** | 0.00 |
| SD05POST-LHS-LOS | 0.56 | **0.83** | 0.73 | 0.78 | **0.87** | 0.82 | 0.00 | **0.00** | 0.00 |
| SD05POST-RHS-CU | 0.84 | **0.81** | 0.77 | 0.76 | **0.65** | 0.76 | 0.01 | **0.02** | 0.01 |
| SD05PRE-LHS-CU | 0.32 | **0.79** | 0.66 | 0.65 | **0.85** | 0.85 | 0.02 | **0.00** | 0.00 |
| SD05PRE-LHS-LOS | 0.8 | **0.7** | 0.65 | 0.85 | **0.91** | 0.73 | 0.00 | **0.00** | 0.01 |
| SD05PRE-RHS-CU | 0.55 | **0.55** | 0.74 | 0.78 | **0.78** | 0.73 | 0.01 | **0.01** | 0.01 |
| SD05PRE-RHS-LOS | 0.53 | **0.57** | 0.7 | 0.61 | **0.78** | 0.84 | 0.03 | **0.01** | 0.00 |
| **Mean** | **0.51** | **0.69** | **0.78** | **0.70** | **0.76** | **0.75** | | | |

Table 3 – Sagittal planes. Overlap coefficients and Kendall's tau rank correlation coefficients with associated p-values demonstrating level of statistical agreement between mass deposition distributions analyzed by sagittal (YZ) planes

| AXIAL PLANES | Overlap Coefficients (OCs) | | | Kendall's tau values | | | p-values | | |
|---|---|---|---|---|---|---|---|---|---|
| **Model** | Neg | **Ref** | Pos | Neg | **Ref** | Pos | Neg | **Ref** | Pos |
| SD02POST-LHS-CU | 0.53 | **0.53** | 0.54 | 0.79 | **0.81** | 0.85 | 0.00 | **0.00** | 0.00 |
| SD02POST-LHS-LOS | 0.41 | **0.5** | 0.54 | 0.63 | **0.67** | 0.75 | 0.00 | **0.00** | 0.00 |
| SD02PRE-LHS-CU | 0.23 | **0.37** | 0.43 | 0.60 | **0.63** | 0.69 | 0.01 | **0.00** | 0.00 |
| SD02PRE-LHS-LOS | 0.26 | **0.33** | 0.44 | 0.69 | **0.69** | 0.71 | 0.00 | **0.00** | 0.00 |
| SD04POST-LHS-CU | 0.56 | **0.6** | 0.63 | 0.76 | **0.83** | 0.83 | 0.00 | **0.00** | 0.00 |
| SD04POST-LHS-LOS | 0.36 | **0.42** | 0.49 | 0.70 | **0.76** | 0.79 | 0.00 | **0.00** | 0.00 |
| SD04POST-RHS-CU | 0.64 | **0.64** | 0.62 | 0.76 | **0.76** | 0.79 | 0.00 | **0.00** | 0.00 |
| SD04PRE-LHS-CU | 0.59 | **0.66** | 0.75 | 0.69 | **0.64** | 0.74 | 0.00 | **0.01** | 0.00 |
| SD04PRE-LHS-LOS | 0.46 | **0.54** | 0.62 | 0.76 | **0.78** | 0.74 | 0.00 | **0.00** | 0.00 |
| SD04PRE-RHS-CU | 0.56 | **0.68** | 0.67 | 0.79 | **0.86** | 0.89 | 0.00 | **0.00** | 0.00 |
| SD04PRE-RHS-LOS | 0.7 | **0.82** | 0.79 | 0.82 | **0.89** | 0.84 | 0.00 | **0.00** | 0.00 |
| SD05POST-LHS-CU | 0.67 | **0.76** | 0.73 | 0.84 | **0.76** | 0.76 | 0.00 | **0.00** | 0.00 |
| SD05POST-LHS-LOS | 0.51 | **0.56** | 0.61 | 0.67 | **0.70** | 0.70 | 0.00 | **0.00** | 0.00 |
| SD05POST-RHS-CU | 0.68 | **0.59** | 0.52 | 0.78 | **0.81** | 0.72 | 0.00 | **0.00** | 0.00 |
| SD05PRE-LHS-CU | 0.6 | **0.74** | 0.67 | 0.85 | **0.85** | 0.86 | 0.00 | **0.00** | 0.00 |
| SD05PRE-LHS-LOS | 0.69 | **0.76** | 0.81 | 0.79 | **0.79** | 0.85 | 0.00 | **0.00** | 0.00 |
| SD05PRE-RHS-CU | 0.65 | **0.72** | 0.67 | 0.72 | **0.72** | 0.72 | 0.00 | **0.00** | 0.00 |
| SD05PRE-RHS-LOS | 0.67 | **0.73** | 0.75 | 0.76 | **0.79** | 0.78 | 0.00 | **0.00** | 0.00 |
| **Mean** | **0.54** | **0.61** | **0.63** | **0.74** | **0.76** | **0.78** | | | |

Table 4 – Axial planes. Overlap coefficients and Kendall's tau rank correlation coefficients with associated p-values demonstrating level of statistical agreement between mass deposition distributions analyzed by axial (XZ) planes

**FIGURES/LEGENDS**

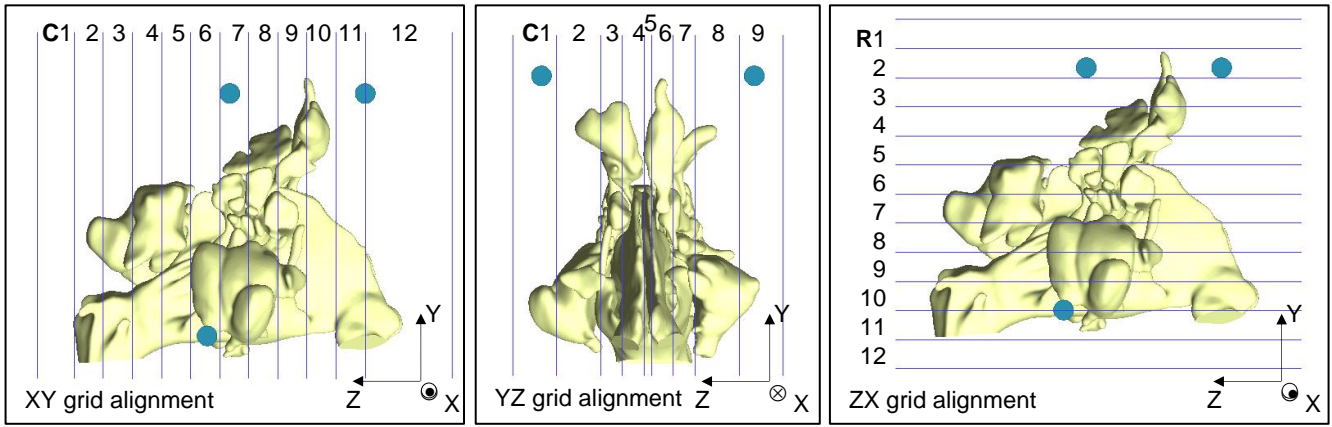

Figure 1 – Grid Creation. Illustrated here is the SD05-PRE (pre-surgical) CFD model with the three orthogonal grid systems designated by columns (C1-12) in the xy (coronal) plane, columns (C1-9) in the yz (sagittal) plane, and rows (R1-12) in the zx (axial) plane.

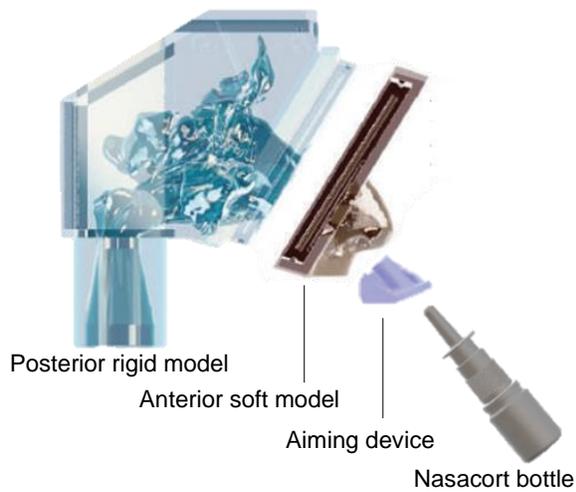

Posterior rigid model
Anterior soft model
Aiming device
Nasacort bottle

Figure 2 – Experimental set-up. Components of experimental set-up are shown in expanded view here for clarity. For each model, Nasacort™ bottle was consistently positioned within the soft portion of the nose using the 3D-printed aiming device. Anterior and posterior portions of the model were printed to fit with snap-on interface. (From Basu et al. (submitted). Used with permission.)[22]

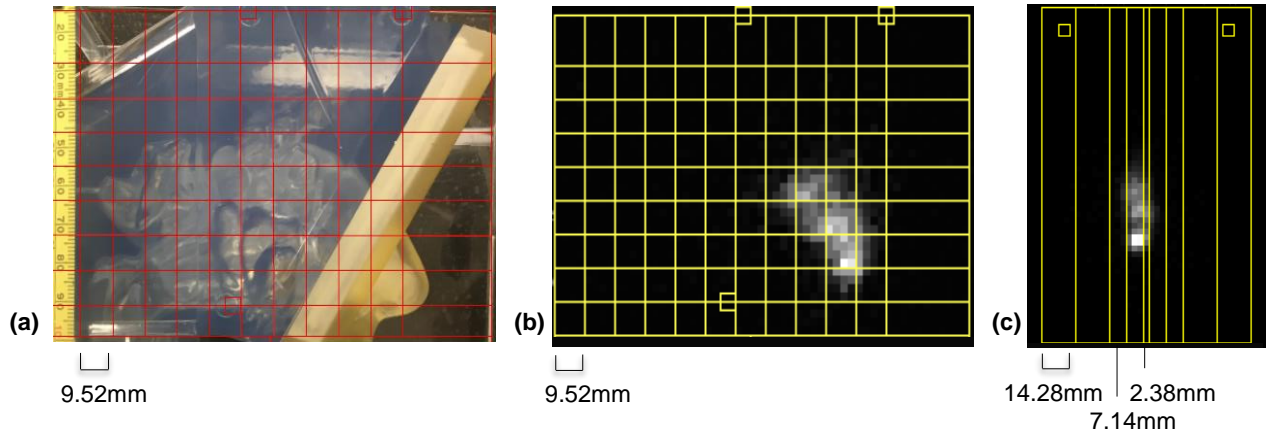

Figure 3 (a) Superimposed grid framework overlying the 3D-printed model of SD05-PRE. (b) Grid superimposed over gamma scintigraphy results in a sagittal view. Smaller boxes mark the positions of americium reference markers. (c) Grid superimposed over gamma scintigraphy results in a frontal view.

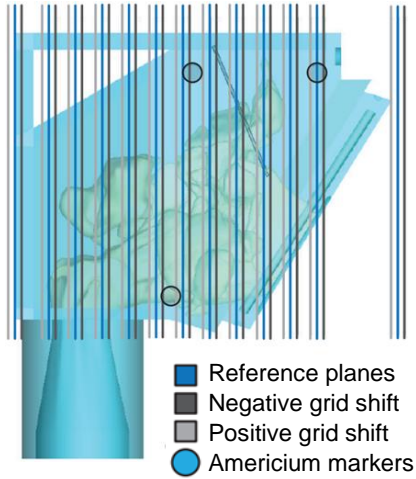

Figure 4 – Grid Shifts. Reference planes were positioned in closest match to the experimental grid system based on the most anteriorly-located americium marker. Reference planes were then shifted by 1 pixel (2.38mm) in the negative and positive direction along the corresponding axis to achieve negative and positive grid shifts, and to account for variation in experimental grid positioning. (From Basu et al. (submitted). Used with permission.)[22]

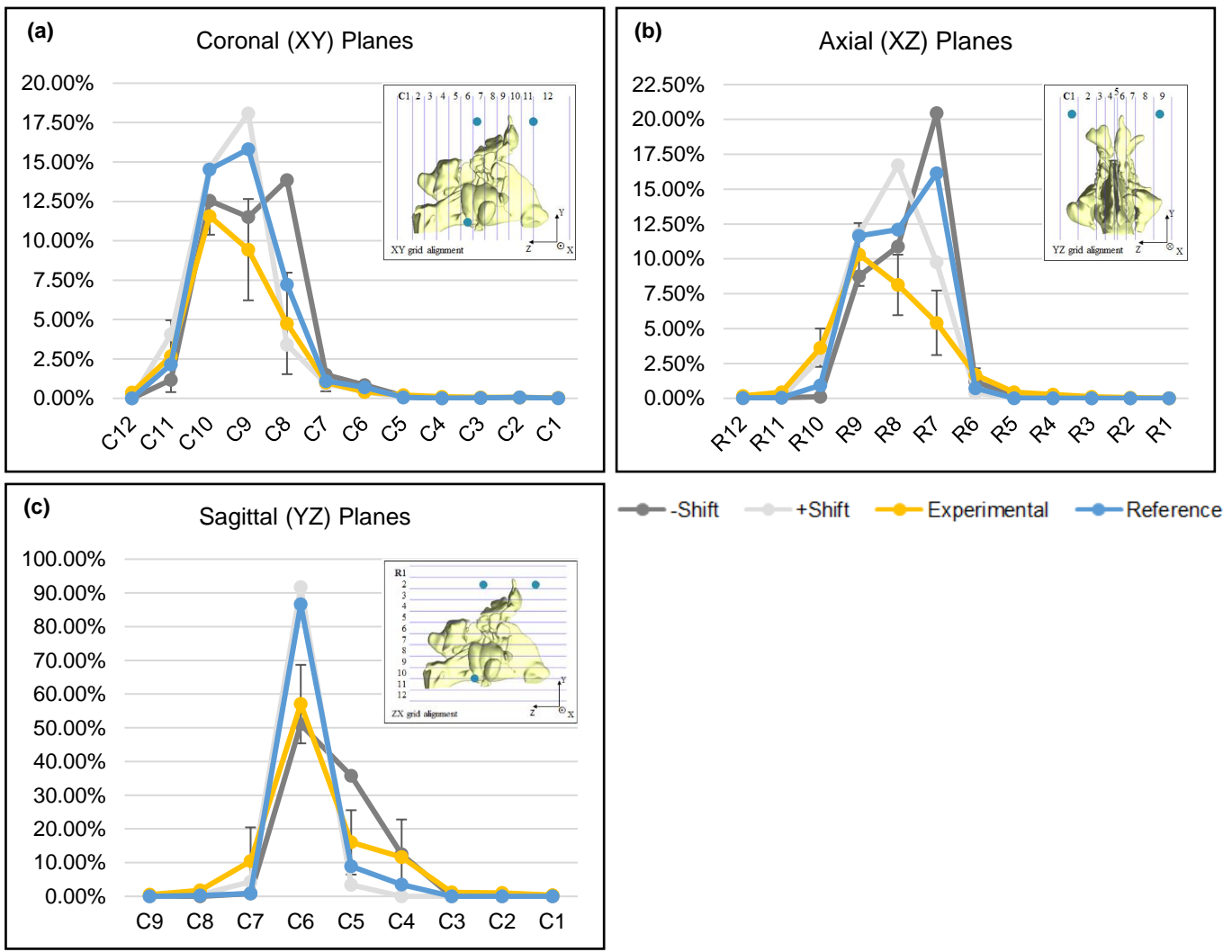

Figure 5 – Fractional mass deposition by grid compartment for Model SD05PRE-LHS-LOS. Inset within each graph is the representative grid system previously illustrated in Figure 1.

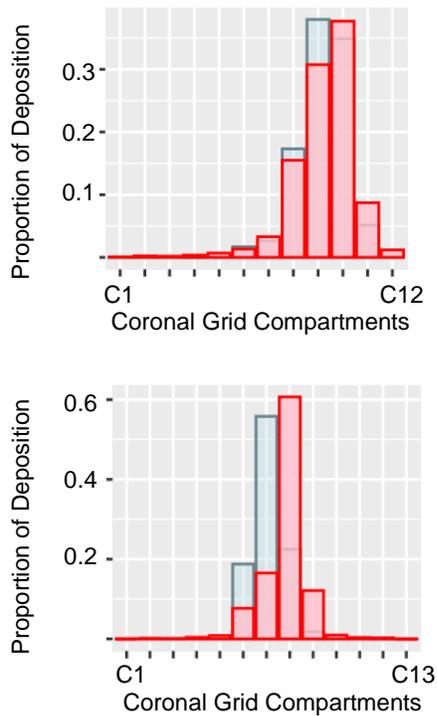

Figure 6 – Overlap coefficients displayed graphically. Figure 7a depicts the strong overlap coefficient (0.91) for Model SD05PRE-LHS-LOS in the reference coronal grid position. Here the computational deposition amounts are depicted in blue and the experimental values in red. The experimental values overlay the simulation so that any overlap is represented as red. The blue portions thus characterize a lack of overlap. Figure 7b depicts the less strong overlap coefficient (0.50) for Model SD02POST-LHS-CU in the negative sagittal grid position. Although there is less agreement between computational and experimental results, the discrepancy occurs in adjacent compartments, maintaining the overall shape of the mass distribution.